\documentclass{aa}  

\usepackage{graphicx}
\usepackage{txfonts}
\usepackage[section]{placeins}
\newcommand{\hd}{HD\,189733}
\newcommand{\xmm}{{\em XMM-Newton}}
\newcommand{\pn}{{\em pn}}
\newcommand{\fxu}{erg s$^{-1}$ cm$^{-2}$}
\newcommand{\lxu}{erg s$^{-1}$}
\begin{document}

\title{X-ray variability of \hd\ across eight years of \xmm\ observations}
\author{I. Pillitteri\inst{1} \and G. Micela\inst{1} \and A. Maggio\inst{1} 
\and S. Sciortino\inst{1} \and J. Lopez-Santiago\inst{2} }

\institute{INAF-Osservatorio Astronomico di Palermo, Piazza del Parlamento 1, 90134 Palermo, Italy\\
 \email{ignazio.pillitteri@inaf.it} 
\and 
Signal Processing Group Dpt. of Signal Theory and Communications, 
Universidad Carlos III de Madrid, Avda. de la Universidad 30 28911 Leganés, Madrid, Spain}

   \date{Received ; }

 
\abstract
{The characterization of exoplanets, their formation, evolution, and chemical changes is tightly linked to our
knowledge of their host stars. In particular, stellar X-rays and UV emission have a strong impact on the dynamical and chemical evolution of planetary atmospheres. 
We analyzed 25 \xmm\ observations encompassing about eight years and totaling about 958 ks in order to 
study the X-ray emission of \hd~A. 
We find that the corona of \hd~A has an average temperature of 0.4 keV and it is only during flares that the mean temperature
increases to 0.9 keV. Apart from the flares, there is no significant change in the flux and hardness of the 
coronal emission on a timescale of several months to years. Thus,
we conclude that there is no detectable activity cycle on such timescales. 
{ We identified the flares and built their energy distribution.
{ The number of flares observed around the phases of the planetary eclipses is not 
statistically different from the number of flares during transit phases. 
However, we do find a hint of a difference in the flare-energy distributions,
as the  flares observed around the planetary eclipses tend to be more energetic than the flares 
observed around the primary transits of the planet.}
We modeled the distribution of the number of flares per day with a power law, showing that it is steeper than the one observed in the Sun and in other Main Sequence stars.
The steepness hints at a significant fraction of undetected micro-flares.
Altogether, the plasma temperatures below 1 keV observed during the flares, along with 
{ the slightly larger} fraction of energetic flares seen at the secondary transits 
highlight the peculiarity of the corona of \hd\,A and points to  star-planet interaction as the plausible origin of part of its X-ray emission. 
However, more observational and modeling efforts are required to confirm or disprove this scenario. }
}

\maketitle
%

\section{Introduction}
{ More than 4500 extra-solar planets are already known thanks to spaceborne missions 
such as Kepler \citep{Kepler} and TESS \citep{TESS} or ground-based photometric and spectroscopic surveys.
However, the fraction of exoplanets that can be characterized with respect to their masses, 
density, and chemical composition is biased towards planets orbiting very close to their hosts. 
Transmission  spectroscopy to infer the chemical composition of the planetary atmospheres 
is more feasible in close-in planets than in exoplanets in wider orbits. 
In the future, the next JWST observations \citep{JWST} and Ariel \citep{Tinetti2018} will enable detailed chemical 
studies of the atmospheres of a significant fraction of known exoplanets.
A full characterization of these atmospheres requires a multi-wavelength approach
to reveal the high energy flux and space weather conditions 
where exoplanets are in orbit.

To properly investigate the conditions for the formation and the evolution 
of planetary atmospheres, the effects that originate from 
close star-planet separations must be taken into account.  
These effects are related to the intense flux originating from 
the star especially at high energies, namely UV and X-rays bands, 
the stellar wind and the magnetized medium where such planets orbit.
Models of photo-chemical evolution (\citealp{Cecchi-Pestellini2009},  
\citealp{Locci2022}) and evaporation \citep{Kubyshkina2018a} 
need to be constrained by the dose of XUV fluxes 
to which the planets are exposed. In this context, a detailed characterization of the stellar coronae, 
their X-ray spectra, and their time variability in form of flares and
coronal mass ejections (CMEs) is of crucial importance. 
The present paper addresses the X-ray characterization of the corona of \hd, 
which is one of the best-studied  exoplanet host. Our goal is to provide the most updated
description of the X-ray emission of this nearby system.

\hd\ is a wide binary system composed of a K1 type star and a M star \citep{Bakos2006}. 
The stellar companion, \hd~B, is an M4 type star in an 3200 yr period orbit at 
about 216 AU of projected separation  from the primary.
\hd A hosts a transiting hot Jupiter in a 2.2 days orbit \citep{Bouchy2005}.  }
Extensive observations in the UV and optical/IR bands have enabled the inference that 
the planet of \hd~A is inflated and actively evaporating \citep{Lecavelier2012,Bourrier2013,Bourrier2020}.

X-rays from \hd A have been investigated in detail with Chandra and \xmm\ 
\citep{Pillitteri2010,Pillitteri2011,Pillitteri2014,Poppenhaeger2014,Pillitteri2015}.
\hd~A has an X-ray luminosity of order of 10$^{28}$ \lxu and a corona with mean temperature 
around 5 MK, thus brighter and hotter than the solar corona.
\hd~B\ has been detected in Chandra observations \citep{Poppenhaeger2013} with a quiescent 
X-ray luminosity of $L_\mathrm X \sim4.7\times10^{26}$ \lxu and a coronal temperature of about 3.5 MK.

The ages of \hd~A and \hd~B derived from their levels of X-ray and chromospheric activity are
discrepant with \hd~A looking younger ($\leq 1$ Gyr)) than \hd~B (few Gyr). \citet{Pillitteri2010} and 
\citet{Poppenhaeger2014} suggested that  the star has been spun up from the exchange of 
angular momentum between the planet and its star.

\hd~A is the first star for which a planetary transit in X-rays{ has been reported} based on the
stacked light curve of six Chandra observations \citep{Poppenhaeger2013}. 
The authors suggested that the radius of \hd~Ab in X-rays is larger than
in the optical band due to the opaque envelope of the inflated atmosphere surrounding the planet.

In this paper, we present a comprehensive analysis of the observations of \hd\ made with \xmm\ 
for a total of $\sim$958 ks. In particular, we study the coronal variability and the occurrence of flares, 
their duration, shape, and energy.
We compare these characteristics to those of other variable coronal sources to seek any difference 
or analogy with flares of single stars without hot Jupiters. 
The structure of the paper is as follows.
In Section \ref{obs}, we present the archival observations and their analysis.
In section \ref{res}, we present our results and 
in Section \ref{conclusions}, we discuss the results and present our conclusions.
\section{Observations and data analysis\label{obs}}
\begin{figure}
\resizebox{\columnwidth}{!}{
  \includegraphics{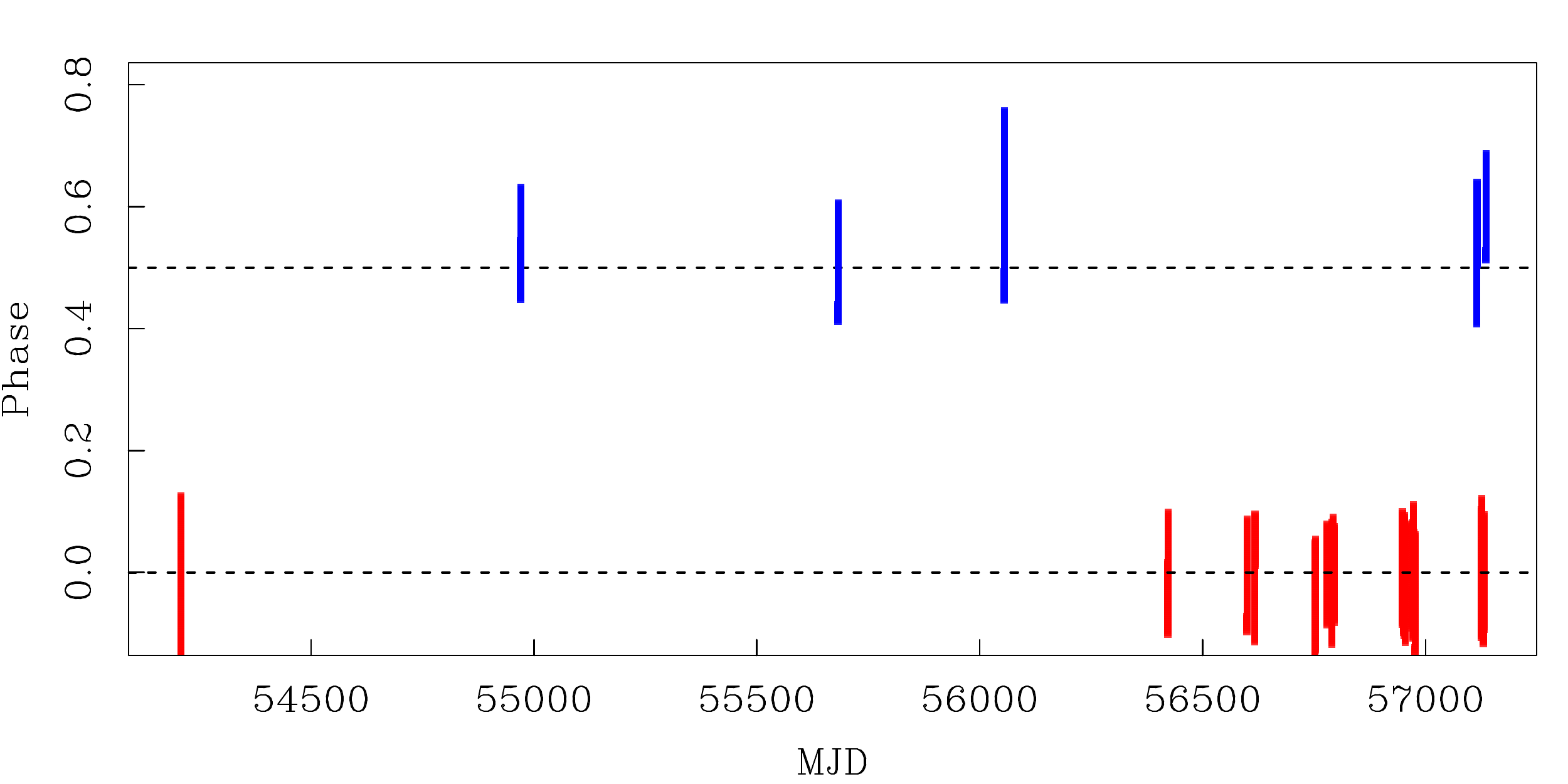}
}
\caption{\label{barlog} 
Planetary phases observed as a function of MJD. 
Phases of the primary and secondary transits are marked by horizontal lines for sake of clarity.}
\end{figure}

\hd\ has been observed 25 times with \xmm\  during primary (phase $\phi=0$) and secondary ($\phi=0.5$) 
planetary transits. 
Table \ref{obslog} lists the available observations and their duration. 
The observations span about eight years between 2007 and 2015 for a total exposure time of 1012 ks 
(with science exposures totaling 958 ks).  
Figure \ref{barlog} shows the range of observed phases  during each exposure.
About 20 out of 25 observations were taken at the primary transits of the planet, 
the remaining five were taken at or just after secondary transits. 
About 400 ks accumulated in 16 observations are part of a large program  (P.I. Wheatley) 
that observed planetary transits of \hd A~b. 
The typical length of the exposures is of order of 35-40 ks, however 
the fourth observation was more than 60 ks long.  
Twenty of the observations were acquired with the {\em Thin1} filter, while five were obtained with the 
{\em Medium} filter. The {\em Thin1} filter has a slightly better sensitivity than the 
{\em Medium} filter at very low energies ($<0.5$ keV), 
however, it could be more prone to high background episodes and UV leaks.

The five observations carried with the {\em Medium} filter were acquired in {\em FullWindow} 
mode (field of view $\sim35 \arcmin$),
while the remaining 20 observations that used the {\em Thin1} 
filter were obtained in {\em SmallWindow} mode, where only a fraction of the chips were used to imaging the 
region around the target. The faster timing allowed by the {\em SmallWindow} mode reduces pile-up effects, however 
this  is irrelevant in the case of \hd~A, since the source count rate is not exceptionally 
high (as compared to other very bright astrophysical sources).
\begin{table*}[!t]
\caption{\label{obslog} Log of the \xmm\ observations. } 
\begin{center}
\resizebox{0.96\textwidth}{!}{
\begin{tabular}{ccccccccc}
  \hline
\hline
 & ObsId & RA\_NOM & Dec\_NOM & Start\_UTC & Duration (ks) & P.I. & Filter & Planetary Phases \\ 
  \hline
   1 & 0506070201 & 20:00:43.70 & +22:42:39.0 & 2007-04-17 14:06:31.000 & 54.9 & Wheatley & Thin1  & 0.84--1.13  \\ 
   2 & 0600970201 & 20:00:43.70 & +22:42:39.0 & 2009-05-18 21:15:54.000 & 37.3 & Wolk & Medium     & 0.45--0.63 \\ 
   3 & 0672390201 & 20:00:43.70 & +22:42:39.0 & 2011-04-30 23:14:20.000 & 39.1 & Pillitteri & Medium &  0.41--0.6 \\ 
   4 & 0690890201 & 20:00:43.70 & +22:42:39.1 & 2012-05-07 18:24:32.000 & 61.5 & Pillitteri & Medium &  0.45--0.76 \\ 
   5 & 0692290201 & 20:00:43.70 & +22 42 35.8 & 2013-05-09 20:16:00.000 & 39.2 & Wheatley & Thin1   & 0.90--1.10 \\ 
   6 & 0692290301 & 20:00:43.71 & +22 42 35.8 & 2013-11-03 07:54:13.000 & 37.1 & Wheatley & Thin1   & 0.90--1.09 \\ 
   7 & 0692290401 & 20:00:43.71 & +22 42 35.8 & 2013-11-21 00:58:40.000 & 42.0 & Wheatley & Thin1   & 0.89--1.09  \\ 
   8 & 0744980201 & 20:00:43.71 & +22 42 35.3 & 2014-04-05 05:05:20.000 & 48.0 & Wheatley & Thin1   & 0.81--1.05 \\ 
   9 & 0744980301 & 20:00:43.71 & +22 42 35.3 & 2014-05-02 01:22:25.000 & 33.7 & Wheatley & Thin1   & 0.91--1.08  \\ 
  10 & 0744980401 & 20:00:43.71 & +22 42 35.3 & 2014-05-13 01:55:22.000 & 40.0 & Wheatley & Thin1   & 0.88--1.08  \\ 
  11 & 0744980501 & 20:00:43.71 & +22 42 35.3 & 2014-05-15 09:57:00.000 & 32.0 & Wheatley & Thin1   & 0.93--1.09 \\ 
  12 & 0744980601 & 20:00:43.71 & +22 42 35.3 & 2014-05-17 14:21:12.000 & 32.0 & Wheatley & Thin1   & 0.92--1.07 \\ 
  13 & 0744980801 & 20:00:43.71 & +22 42 35.3 & 2014-10-17 16:08:26.000 & 37.0 & Wheatley & Thin1   & 0.87--1.06 \\ 
  14 & 0744980901 & 20:00:43.71 & +22 42 35.3 & 2014-10-19 20:38:36.000 & 34.4 & Wheatley & Thin1   & 0.91--1.1 \\ 
  15 & 0744981001 & 20:00:43.71 & +22 42 35.3 & 2014-10-22 01:39:14.000 & 39.9 & Wheatley & Thin1   & 0.9--1.07 \\ 
  16 & 0744981101 & 20:00:43.71 & +22 42 35.3 & 2014-10-24 06:15:47.000 & 39.0 & Wheatley & Thin1   & 0.9--1.09 \\ 
  17 & 0744981301 & 20:00:43.71 & +22 42 35.3 & 2014-11-08 20:16:34.000 & 34.6 & Wheatley & Thin1   & 0.88--1.08 \\ 
  18 & 0744981201 & 20:00:43.71 & +22 42 35.3 & 2014-11-11 00:37:26.000 & 44.0 & Wheatley & Thin1   & 0.89--1.11 \\ 
  19 & 0744981401 & 20:00:43.71 & +22 42 35.3 & 2014-11-13 06:46:05.000 & 32.0 & Wheatley & Thin1   & 0.91--1.08 \\ 
  20 & 0744980701 & 20:00:43.71 & +22 42 35.3 & 2014-11-15 09:48:00.000 & 39.0 & Wheatley & Thin1   & 0.91--1.06 \\ 
  21 & 0744981501 & 20:00:43.71 & +22 42 35.3 & 2015-04-13 02:37:23.000 & 45.4 & Wheatley & Thin1   & 0.89--1.12 \\ 
  22 & 0744981601 & 20:00:43.71 & +22 42 35.3 & 2015-04-17 12:34:55.000 & 41.0 & Wheatley & Thin1   & 0.88--1.09 \\ 
  23 & 0744981701 & 20:00:43.71 & +22 42 35.3 & 2015-04-19 19:06:26.000 & 38.0 & Wheatley & Thin1   & 0.91--1.09 \\ 
  24 & 0748391401 & 20:00:43.69 & +22:42:39.1 & 2015-04-03 03:30:49.000 & 47.4 & SCHARTEL (PS) & Medium &  0.41--0.64 \\ 
  25 & 0748391501 & 20:00:43.69 & +22:42:39.1 & 2015-04-23 06:17:15.000 & 44.0 & SCHARTEL (PS) & Medium &  0.51--0.69 \\ 
   \hline
\end{tabular}
}
\end{center}
Note: The observations are listed by increasing ObsID.
The range of orbital phases of the planet are also listed.
The last two observations were obtained with a request of Director Discretionary Time (DDT, P.I. Schartel).
\end{table*}
 
The datasets were downloaded from the \xmm\ web archive\footnote{\url http://nxsa.esac.esa.int/nxsa-web/#search} 
and reduced with SAS v18.0 in order to obtain tables of events
calibrated in sky and detector positions, arrival times, pulse energies, and quality grades. 
The selection of events and their filtering in appropriate energy bands and quality grade has been performed through
the SAS task {\sc evselect}. 
The same task allows us to accumulate light curves and spectra for the EPIC datasets.

\subsection{Contamination from nearby sources}
\begin{figure}
\resizebox{\columnwidth}{!}{
  \includegraphics{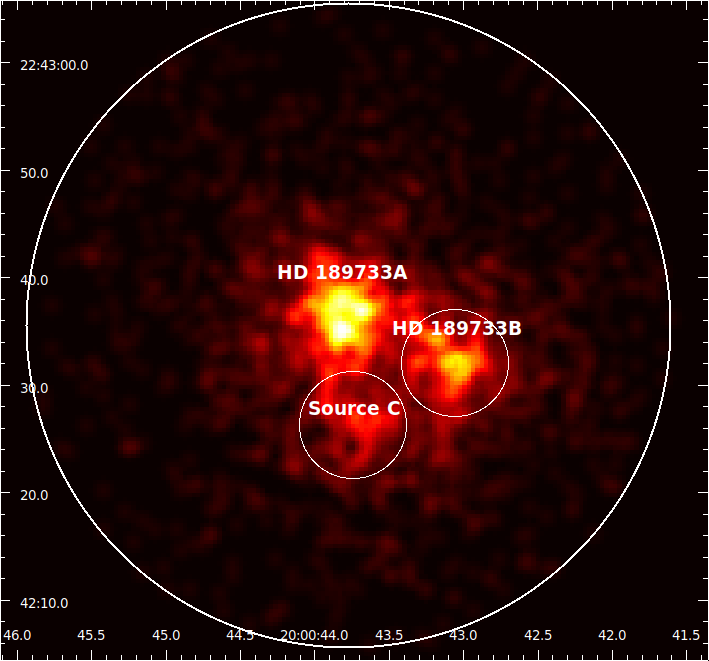}
}
\caption{\label{img} MOS1 image of \hd\ during observation 0744981201. The brightest source is \hd~A,
followed by  \hd~B and source C (likely a background AGN). 
The circles indicate regions of size 5\arcsec. The large circle
has radius 30\arcsec. \hd~B was flaring during this observation while during quiescence was barely detected.}
\end{figure}

A circular region of a 30\arcsec  radius used to extract the events associated with an X-ray source imaged with \xmm-\pn\ 
contains about 80\% of the total events emitted by the source. 
\hd\ has two nearby X-ray sources within about 12\arcsec: the M type star \hd~B, 
and an X-ray source that is likely of extra-galactic nature and perhaps associated with an AGN (hereafter source C)
\citep{Pillitteri2010,Poppenhaeger2013}. Figure \ref{img} shows the positions in one of the MOS1 images.
We display here the MOS because allows for a better resolution over the \pn. For the analysis of light curves and spectra,
we used the data from \pn\ because they offer a higher count statistics over MOS.
The contribution of events from these sources to the light curves and spectra of \hd~A must thus be taken into account.
For \hd~B and source C we extracted the events and light curves in two 5\arcsec regions around their centroids. 
Source C has a hard spectrum \citep{Pillitteri2010,Pillitteri2011} and, from an examination of its light curves, 
it does not seem to vary significantly in time.
These two conditions allow us to minimize the contamination from source C by simply restricting our analysis to the
band $0.3-2.0$ keV and excising the core of the counts centroid up to 8\arcsec. 
Previous analyses of the X-ray spectra of \hd~A have demonstrated that most of the coronal
emission of \hd~A is concentrated at energies below 1 keV \citep{Pillitteri2014}.

\begin{figure*}
\centering
\resizebox{0.8\textwidth}{!}{
  \includegraphics[height=8cm]{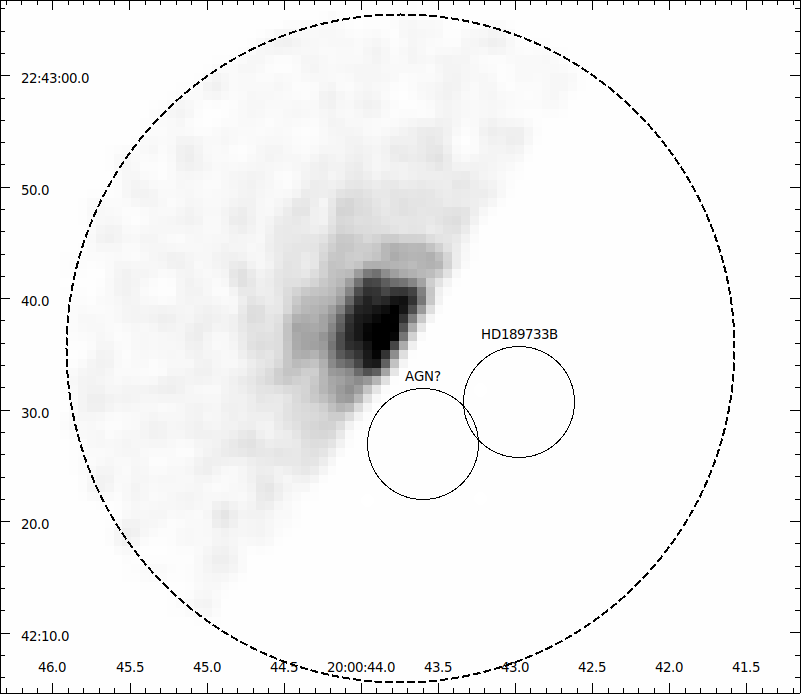}
  \includegraphics[height=8.2cm]{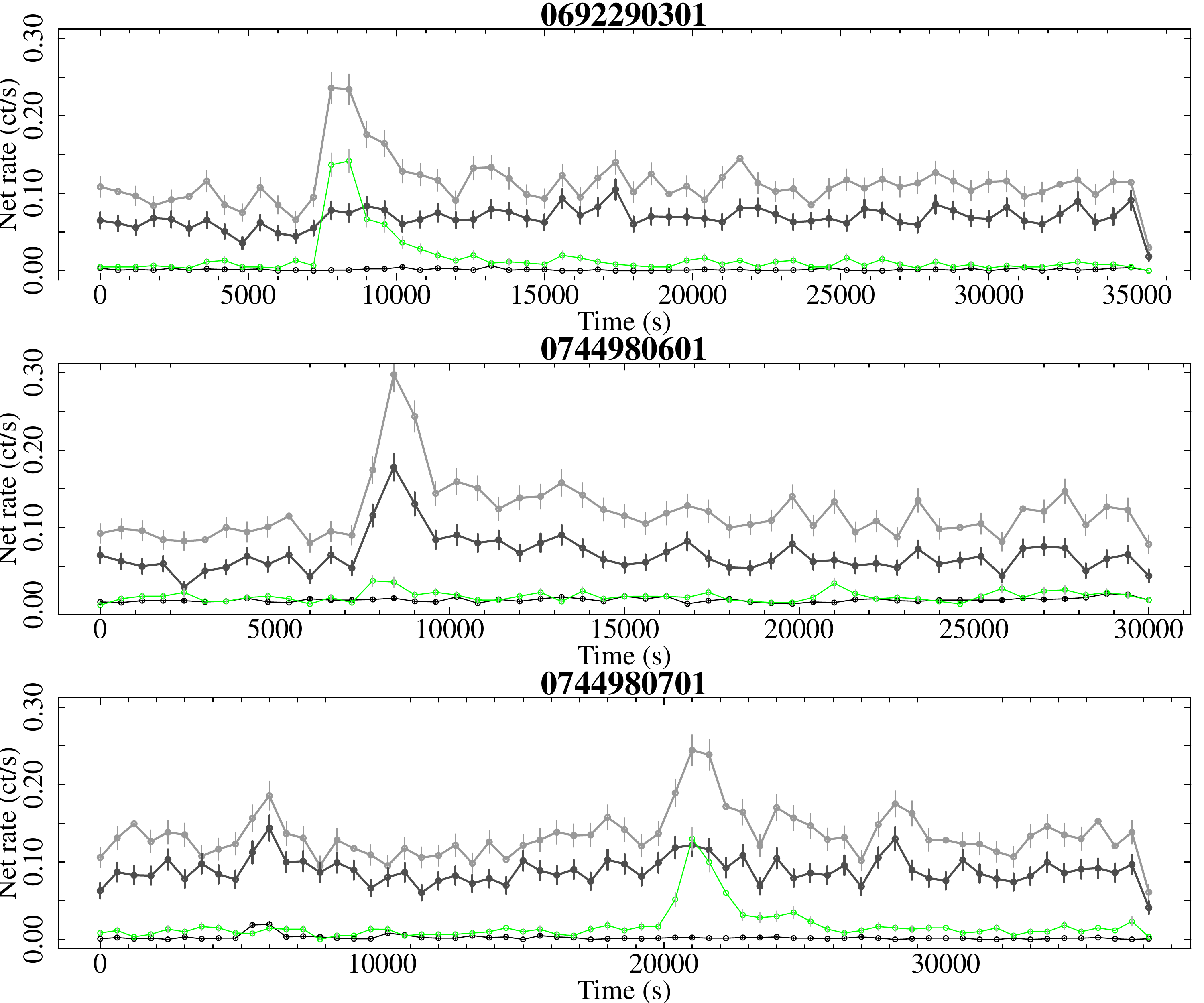}
}
\caption{\label{testlc} { Discriminating among different sources of variability.}
Left panel:  \pn\ image and selection of events of \hd~A in  
a semicircular region during observation 0744981201 in the band 0.3-2.0 keV. 
The circles indicate regions of size 5\arcsec. The large circle
has radius 30\arcsec. Right panel: Three examples of light curves where flaring of \hd~A or \hd~B is detected.
The gray light curves are  from events in a 30\arcsec region minus two 8\arcsec regions around \hd~B and source C
(selection {\em i + ii} ). The time bins are 600 s each.
The black light curves are accumulated from the events in the semicircular region opposite to \hd~B and source C 
as the figure in the left panel. 
Green light curves are made from the events of \hd~B in two 5\arcsec regions. 
The background light curves scaled to the area of selection {\em i + ii} are also shown.
In the top panel, a flare is visible both in the gray and green curves, but not in the black curve; 
this allows us to conclude that the flaring source was \hd~B and contaminated the selection in {\em i + ii} 
but not the selection in the semicircular region on the opposite side. 
In the middle panel, a flare is detected both in gray and black curves,
this flare is genuinely due to \hd~A. 
In the bottom panel three flares are detected, the first one, seen in gray and black curves,  attributed to \hd~A; while the second one is visible in gray and green curves and attributed to \hd~B.
Another small flare of \hd~A is visible after the flare of \hd~B.
}
\end{figure*}

The operative way to minimize the contamination of the light curves of \hd~A is thus summarized as follows:
\begin{itemize}
\item {\em i)} select \pn\ events in a 30\arcsec region centered on \hd~A and in the band 0.3-2.0 keV;
\item {\em ii)} exclude the events in two 8\arcsec circular regions centered on the positions of \hd~B and source C;
\item {\em iii)}  select the events in a semi-circular region opposite;
to and not encompassing the positions of \hd~B and source C. This is a control set for the variability of \hd~B, 
(cf. Fig. \ref{testlc});
\item {\em iv)} select events from a 40\arcsec circular region for background subtraction to be applied to the light curves 
 obtained in {\em ii} and {\em iii} 
\end{itemize}

{ Light curves with binning of 600 s obtained from events} selected at points {\em i} and {\em ii} 
have been used for the analysis of variability  of \hd~A and \hd~B, respectively.
{ The choice of the binning interval is justified by the subsequent use of an algorithm 
to detect variability and flares, which assumes that in each bin, the counts 
follow an underlying Gaussian distribution (see Sect. \ref{flares}).
A bin of 600 s gives on average about 100--150 counts per bin, which fulfills 
the requirement of Gaussianity for all the observations, even during quiescent phases.}
When \hd~B was quiescent, its flux was very low and similar to the local background flux. 
However,  in the case of flares from \hd~B, the light curves of \hd~A obtained at {\em i+ii} 
also show an excess due to the flares from \hd~B. 
The flux of \hd~B can overwhelm the flux of \hd~A at the position of the centroid of 
\hd~A even when excising the events in a 8\arcsec region as in {\em ii}.
Hence, the flares of \hd~B can be mistakenly attributed to \hd~A. 
To check any flaring of \hd~B we used the light curves obtained in point {\em iii} in order to determine the intervals 
with flares from \hd~B in the light curves obtained from points {\em i+ii}.
The light curves of the events in a semicircular
region of 30\arcsec not comprising  \hd~B and source C (point {\em iii}) served as a control sample 
to be compared to the light curves obtained at points {\em i} and {\em ii.} 
Figure \ref{testlc} shows both the selection region in {\em i} through {\em iii, }  along with three examples 
where the light curves of the primary are contaminated by flares of the secondary. 
  
While the background can be variable in our \xmm\ observations, we did not remove high background intervals. We preferred to
keep uninterrupted time series within each observation and subtract the scaled background count rates to 
obtain the net source count rates. { However, we did not consider the intervals of strong background variability 
for the subsequent analysis of flares (see Sect. \ref{res}).}

\section{Results \label{res}}
Here, we introduce the variability analysis of \hd~B and the variability of \hd~A.
\subsection{Flares of the M star \hd~B}
As mentioned in the introduction, \hd~B is a M4 star with a relatively low level of quiescent emission,
We identified the flares by eye as the main episodes of variability, characterized by a steep rise followed by
a slower decay.  We thus identified four bright flares from \hd~B in the 25 \xmm\ observations.
In the remaining 21 observations, \hd~B was quiescent and its flux was barely in excess with respect 
to the background level. 

\begin{figure}
\resizebox{\columnwidth}{!}{
  \includegraphics{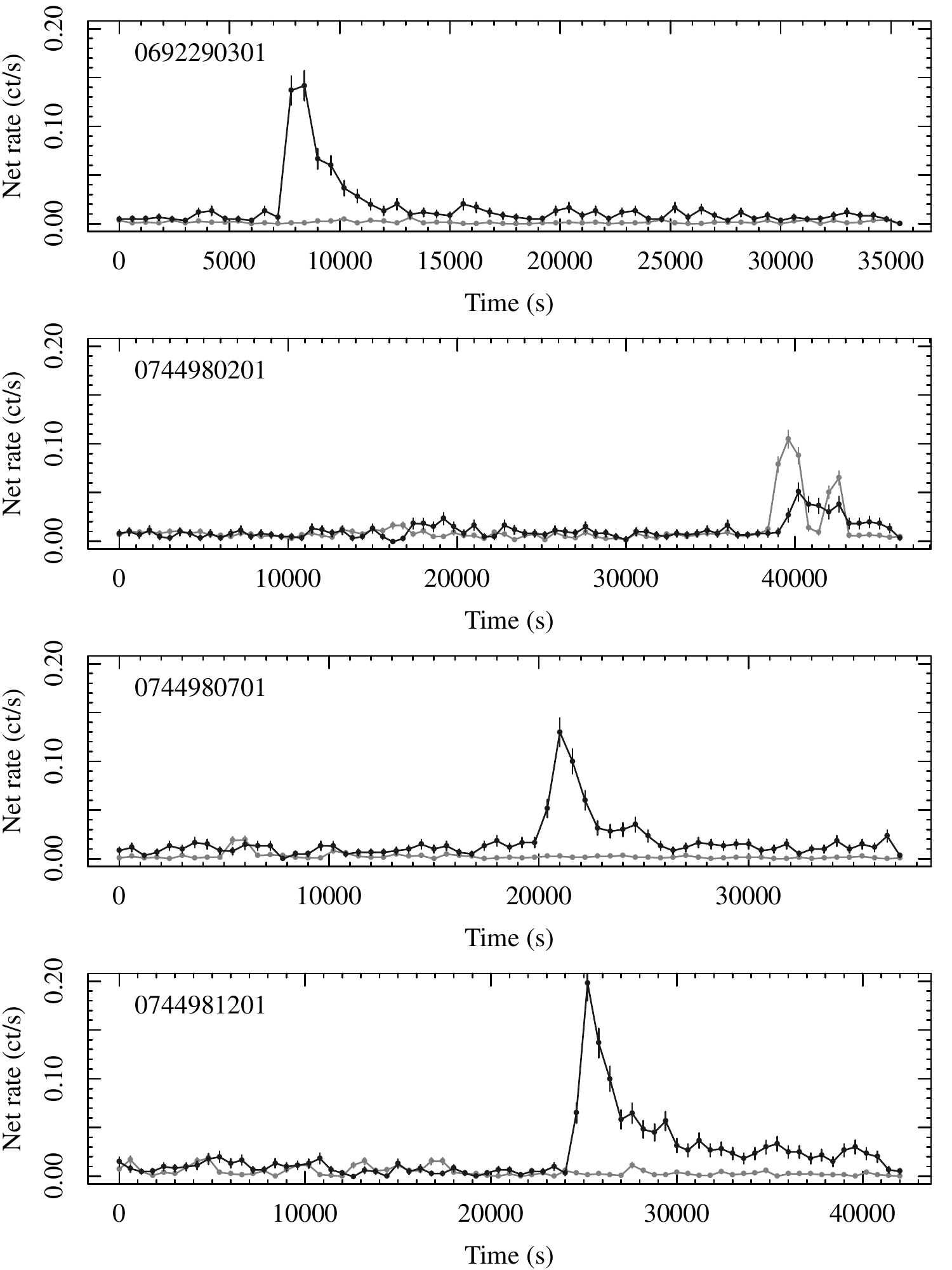}
}
\caption{\label{hdbflares} 
\pn\ light curves of \hd~B with flares. The time bins are 600 s. The light curves of the star (black curve)
 and the scaled background (gray curve) are shown.
}
\end{figure}

Figure \ref{hdbflares} shows the four \pn\ light curves during which \hd~B was flaring. 
The flare in ObsId 0744980201 is heavily affected by background variability, so we do not discuss it in detail here. 
Its duration was about 5 ks with a rise time of about $900-1200$ s, the peak intensity was about a factor 
of four with respect to the quiescent rate, and it is the faintest among the flares detected in \hd~B.

During  observation 0744981201, we observed the brightest flare from \hd~B, with a ratio of the peak-to-
quiescent rate equal to $\sim30$.  The rise phase of the flare was about $1200\pm 300$ s. 
The decay phase is characterized by an initial steep decrease of duration of about
8000 s (e-folding time $\sim3500$ s), followed by a much flatter decay that covers the remaining 
exposure time (about 16 ks). 
The star did not return to its initial quiescent rate, rather it kept emitting a flux about 3.5 times higher than 
the quiescent flux before the flare. Small impulses are visible during this post-flare phase -- 
this flickering perhaps  occurred in the same coronal region after the main ignition resulting 
thus in an overall flux higher than before the flare.

Flares detected in ObsIds 0692290301 and 0744980701 are very similar, with rise times $t_r\sim1200-1500$ s 
and decay encompassing about 5000 s. 
The flare in ObsId 0692290301 has been cited in \citet{Bourrier2020}.
These two flares  were less bright than the flare in 0744981201, reaching a peak rate of 28 and 13 times 
the quiescent rate, respectively. All the three flares seem to show a secondary peak after the initial fast decay 
hinting a second ignition or a triggered ignition in nearby coronal loops after the first event.

The spectra of the quiescent phases are well described with a thermal component at $0.3\pm0.1$ keV with a flux of 
$4-6\times10^{-14}$ \fxu\ in $0.3-2.0$ keV band corresponding to a luminosity of $2-3\times10^{27}$ \lxu. 
These values are a factor of four to five higher than the X-ray luminosity derived by \citet{Poppenhaeger2013} with Chandra.
During the flares a second thermal component is found at $1.3-1.5$ keV
in addition to the cool one. The flux rises by a factor of 10 or more and the luminosity reaches $\sim3\times10^{28}$ \lxu.
Using some diagnostics based on peak temperatures and decay time \citep{Serio1991,Reale2007}, 
we can infer flaring loop sizes of order of 
$5\times10^{10}$ cm, which is comparable to the radius of the star. 
The shape and temperatures of these flares resemble those observed in the Sun and other solar analogs that are 
due to mechanisms of coronal heating and are related to magnetic reconnection. 
The energy released by these flares is of order of $1.2-1.4\times10^{32}$ erg and 
the cumulative energy is more than $4\times10^{32}$ erg released on a total exposure time of 958 ks. 
Two  of these flares were observed four days apart (ObsId 0744980701 and 0744981201). 
We speculate that they could have been originated in the same region -- one that is more active than the rest of the corona.

\subsection{Emission and variability of \hd~A}
\subsubsection{Hardness ratio and mean coronal temperature of \hd~A} \label{hrsect}
\begin{figure}
\resizebox{\columnwidth}{!}{
  \includegraphics{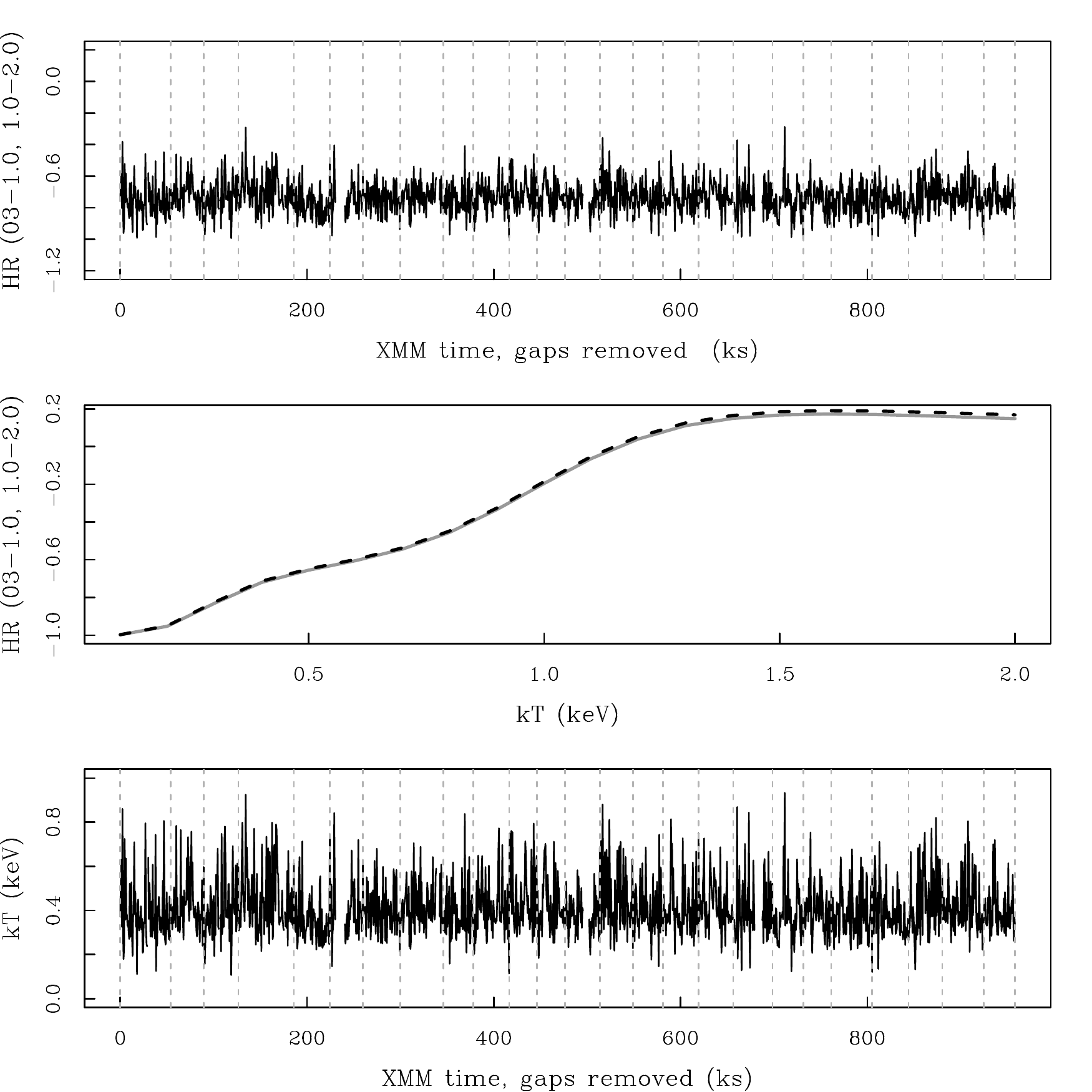}
}
\caption{\label{hr_kt} { Hardness ratio and coronal temperature.} 
Top panel: Time series of the hardness ratio (HR) calculated in the bands $0.3-1.0$ keV (soft) and
$1.0-2.0$ keV (hard) as a function of the exposure time (bins of 600 s). 
Vertical lines indicate the durations of each \xmm\
observation. The average uncertainty of HR is about 0.09 (quantile range 25\%--75\%: 0.08--0.11).
 Middle panel: Relationship between HR and temperature (kT) derived with a library of synthetic
spectra and XSPEC software for the case of {\em Thin1} filter (solid line) and {\em Medium} filter (dashed line). 
Bottom panel: Time series of kT expressed in keV 
(average uncertainty is 0.09 keV, range 25\%--75\%: 0.07--0.12 keV). }
\end{figure}

We constructed the light curves of \hd~A in 0.3-1.0 keV and 1.0-2.0 keV bands with 
binnings of 600 s in order to calculate
an hardness ratio (HR: $(H-S)/(H+S)$) (Fig. \ref{hrpn}). From the HR values we derived a mean coronal temperature
of the star by means of a library of synthetic spectra calculated with {\sc XSPEC} software. The spectra were modeled
with a single thermal APEC component with temperatures in a range between 0.1 and 2.0 keV. From the
synthetic spectra convolved with the spectral response and effective area of \xmm\ \pn,\ we derived counts and HRs
in the bands $0.3-1.0$ keV (soft) and $1.0-2.0$ keV (hard) as for the real data. We used the  
effective area files calculated for the {\em Thin1} and the {\em Medium} filters and used the appropriate set of 
simulated spectra for each observation to calculate the expected HR values corresponding to each value of kT of the
spectra. 
Then we interpolated the simulated HRs onto the observed HRs to obtain the corresponding values of kT. 
Figure \ref{hr_kt} shows the values of HR during the 25 observations as well as the relationship between HR and kT 
of the model and the values of temperatures as a function of the \xmm\ exposure time.  
The differences between the {\em Thin1} and {\em Medium} filters are minimal (less than 1\%).
The intervals where the flares of \hd~B occurred were removed from the analysis.
The kT light curve indicates that almost all the coronal emission of \hd~A is always below 1 keV.
The relationship between HR and kT becomes insensitive for temperature above 1.4 keV
but it works well for \hd~A because of the softness of its spectrum. 
Overall, we do not recognize cyclic variations of the hardness of the coronal spectrum of \hd~A 
over a timescale of eight years or fewer, which could arise from an activity cycle on such a timescale.

\begin{figure}
\resizebox{\columnwidth}{!}{
  \includegraphics{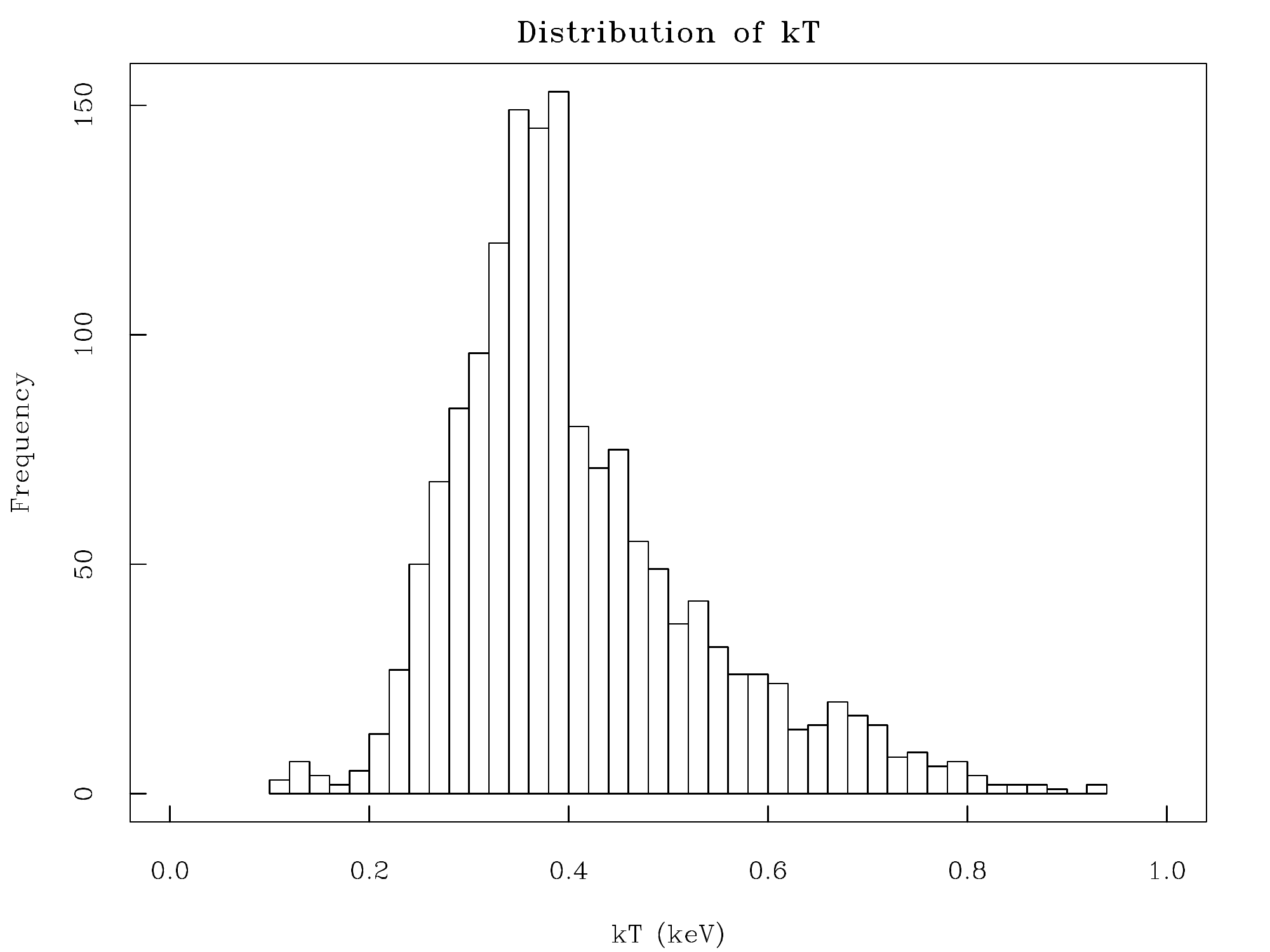}
}
\caption{\label{edist} Histogram  of the values of the mean temperature (kT) inferred from the hardness ratio values. 
}
\end{figure}

Figure \ref{edist} shows the histogram of the values of the mean temperatures (kT)
of the corona of \hd~A derived from HR. 
The distribution of kT has a median of 0.4 keV corresponding to about 4.4 MK, the $25\%-75\%$  quantile range 
is in $0.3-0.5$ keV and the maximum temperature is $\sim0.9$ keV (10.8 MK).

{
From the count rates and the mean coronal kT derived from HR,  we obtained fluxes and luminosities 
by adopting a conversion factor (CF) between count rates and fluxes in the 0.3--2.0 keV band. 
We used PIMMS to convert \pn\ {\it Thin1} ({\it Medium}) filter count rates
to fluxes with a grid of optically thin plasma thermal APEC models spanning kT in 0.1--1.2 keV.
We set N$_H=0$ cm$^{-2}$ as this is compatible with the gas absorption measured from the spectra of \hd~A 
and its nearby distance \citep{Pillitteri2010}, and the overall metal abundance was chosen equal 
{ to $Z=Z\odot$.\footnote{ This choice is supported by the results of the best fit of the full 
exposure spectrum of observation nr. 15 where there was negligible flaring activity. 
From a best fit to a model with two thermal APEC components we obtained (90\% confidence ranges quoted 
in braces): $kT_1=0.25\ (0.22-0.28)$ keV, $kT_2 =0.8\ (0.76-0.91)$ keV, and  $Z=0.15 Z_\odot\ (0.11-0.19)$.} }
We thus derived a relationship between kT and CF for observations made with either {\em Thin1} or 
{\em Medium} filters.
The CF values range between $1.08\times10^{-12}$  and $1.31\times10^{-12}$ erg cnt$^{-1}$ cm$^{-2}$. 
We interpolated these relationships 
to obtain the CFs for each time bin of each observation and relative to {\em Thin1} or {\em Medium} filter. 
We also corrected the count rates for systematic offsets like the fraction of PSF comprised in
the extraction region of the events and dead time due to the read out of the frames with the SAS task {\em epiclcorr}.
From the fluxes, we finally obtained luminosities, assuming a distance of 19.775 pc 
(plx$ = 50.57\pm0.03$ mas, \citealp{GAIAeDR3}).  A synoptic view of the X-ray luminosity 
for all the \xmm\ observations is given in Fig. \ref{lxall}.
}

\subsubsection{Flares of \hd~A}
\label{flares}
Once we identified the main flares of \hd~B, we went on to identify the flares of \hd~A. 
We first discarded the four time intervals where \hd~B was flaring and thus 
the predominant source of variability in the light curves.
For identifying the flares of \hd~A and their lengths, we adopted a semi automatic method. 
We first determined a quiescent rate for each light curve corresponding to the 10\% quantile of the distribution 
of the count rates in that curve. 
This approach excludes the bins where the count rate drops to low values due to telemetry issues 
(as in observation 2 of Fig. \ref{lc1} at $\sim23$ ks).
The median quiescent luminosity of \hd~A in the $0.3-2.0$ keV band is about $4.5\times10^{27}$ \lxu
($25\%-75\%$ range $4.3-5.3\times10^{27}$ \lxu).

{ We used the  {\it ChangePoint} package (version 2.2.2)\ within R language \citep{Killick2011}
to detect changes of the count rates and to segment the light curves in intervals of 
statistically constant count rates. 
In our case we applied the {\em ChangePoint} package with the PELT algorithm to the background subtracted 
count rates of \hd~A and to the light curves of the background. 
The penalty parameter ($p$) is crucial for a coarser or finer segmentation of the
light curve and needs to be properly set. 
A higher $p$ results in fewer time segments, while a lower $p$ produces more segments. 
The optimal value of $p=0.006$ was assessed with a set of simulations of a constant source
with count rates normally distributed around a mean value ($r_q$), equal to the quiescent count rates 
that we determined in each light curve 
and with standard deviation ($\sigma_q$) equal to the median of the errors of the net count rates 
(after propagating the errors due to background subtraction). 
Thus, a value of $p$ was chosen  such that {\em ChangePoint} detects at most one break  
per light curve on average under the expectation that a constant (noisy) count rate produces 
only one time segment equal to the full exposure. 

{ As explained in Sect. \ref{obs}, the bin size of light curves was chosen in a way
to guarantee enough counts per bin. We tested the use of smaller bin sizes of 100 s and 300 s
with {\em Changepoint}. The results were that such finer binning produced a few more short time 
segments  due to noise without additional information on the variability. }

Using simulated light curves, we quantified also the sensitivity to small flares 
that could be confused with small statistical variability and were thus not detected.
To do so, we injected a number of flares in the light curves of constant source described above.
The flare profile was modeled with two exponentials for the rise and the decay phases.
The analytical form is:
\[ F(t) =
  \begin{cases}
    P\,e^{(t-t_0)/\tau_r}       & \quad \mathrm{for\ t <t_0}\\
     P\,e^{-(t-t_0)/\tau_d}      & \quad \mathrm{for\ t >t_0,}
  \end{cases}
\]
where t$_0$ sets the time of the peak, $\tau_r$ and $\tau_d$ are the typical rise and decay times, 
respectively, and P is the peak normalization.
To reduce the parameter space, we chose 1 ks for the rise time and 5 ks 
for the decay time. These values are reasonable choices that match well the observed
characteristic rise and decay phases. The only free parameters of the flare model were thus  the
time of the peak ($t_0$) and intensity of the peak ($P$) in units of sigma ($\sigma_q$). 

{ The number of flares to inject was determined from a power law relationship 
$dN/dE/day \propto E^{-2}$ (see Sect. \ref{conclusions}). 
We simulated flares with energies between  $2\times10^{30}-1.2\times10^{31}$ erg, 
which is the range where we observe a flattening of the flare distribution 
(cf. Fig. \ref{logE}) and  that corresponds to peaks in the range $0.5-1\sigma_q$ 
above the quiescent rates. The expected number of flares per day in this range of energy 
is about 16. We scaled this number for the duration of each observation and, thus, the number of 
injected flares was 5--11, depending on the length of each light curve. 
The {\em ChangePoint} algorithm applied to the simulated light curves resulted in about 
three time segments per light curve due to small flares detected at the penalty  
threshold chosen for the real data. 
These elevated segments are generated from a single flare or, more frequently, from the
blend of several small flares. 
}

Once we identified the time segments with {\em ChangePoint}, we grouped them into
flares defined as the union of contiguous segments starting with a rising or a peaking segment and
followed by one or more decay segments. In observation no. 2, the dip during the decay
is due to a telemetry glitch; in this case, we considered the decay lasting from peak to the
end of the observation. A reference level is given by the quiescent rate, $r_q$, defined above. 
In some cases, the average rate is significantly above the quiescent level in a 
block more than 5 ks long, but it cannot be clearly associated to a single flare. 
In these cases, we checked whether the flickering could be due to blended small flares, 
as has been observed in several simulations. 
In those cases, and only when the average count rate in one block 
was higher than 1$\sigma_q$ above the quiescent level $r_q$,  we 
tried to identify small flares subdividing the elevated interval into small ones. 
As a guidance, we checked the HR that tends to increase during the flares 
because of plasma heating.
We applied the {\em ChangePoint} algorithm to the background curves
as well and we  discarded the time intervals  when the background was 
very high from searching for flares (see Fig.  \ref{lc1}) 

The number of flares identified in this way was 60 -- with 47 occurring in observations taken around
primary transits of \hd~Ab and another 13 occurring in observations taken at secondary transits.
The rates of flares per ks were very similar: $0.063\pm0.008$ ks$^{-1}$ for the total sample, 
$0.065\pm0.01$ ks$^{-1}$ for the observations at primary transits and 
$0.054\pm0.015$ ks$^{-1}$ for the observations at secondary transits.
The rates of flares per day were $5.4\pm0.7 d^{-1}$, $5.7\pm0.8 d^{-1}$, $4.7\pm1.3$, respectively, for the
full sample and the two sub-samples observed at the primary and secondary transits. 

We multiplied the flares luminosities for each bin time (600 s) after subtracting the quiescent level
and summed the energies per bin to obtain the total flare energies.
In the next section, we discuss the properties of the flares, their energies, and the comparison between them and
the flares observed in the Sun and solar type stars.

{ During transits, the planet could obscure a surface spot onto the star with projected 
size similar to that of the planet. Indeed, the available TESS light curve shows that 
some of the profiles of the optical transits are distorted due to the planet passing over a stellar spot 
(Colombo, private communication). 
Moreover, it seems that the same spot lasted for more than one orbital period ($\sim2.2$ days),
so that the planet could have passed in front of the same region more than once.
Photospheric spots are the anchors of magnetic loops and coronal active regions. 
The time required for the planet to transit across an active region of the size of Jupiter is 
about 1000 s. The possibility that the obscured region would flare during that interval and that 
the same flare would be unnoticed is, however, reputed to be low. 
Given the count statistics of the X-ray light curves and the chosen bin time, 
the event is hardly  detectable with a high significance. Moreover, 
the patchy and variable stellar surface in X-rays makes these kinds of obscuration 
even less detectable. }

\section{Discussion and conclusions\label{conclusions}}
In this paper, we analyze, in a homogeneous way, 
25 observations of the \hd\ system obtained with \xmm\ between the years 2007-2015.
This set of observations is quite unique in the context of stars hosting planets because they 
accumulated almost 1 Ms, devoted to observations of the same star in X-ray at two opposite phases of the orbit of its planet.
The present study offers thus a comprehensive characterization of the variability across the lifetime of eight years for \hd~A.

In Table \ref{fltab}, we report the list of identified flares with their duration, energy, total luminosity, and flux.
We also report the flux received at the surface of \hd~Ab.
When considering all the flares, the total flaring time was 423.6 ks out of 958 ks for a
flare percentage of 44\% of the exposure time. 
The total time of the flares that occurred around the secondary transits amounts to about 123.6 ks 
or $\approx51.5$\% of the observing time (240 ks).
The corresponding time at the primary transits was 300 ks or 41.8\% of the observing time (718 ks).
In the observed range of flare energies, the flares detected at secondary transits are more frequent than the
flares detected at primary transits. 

\begin{figure}
\begin{center}
\resizebox{\columnwidth}{!}{
\includegraphics{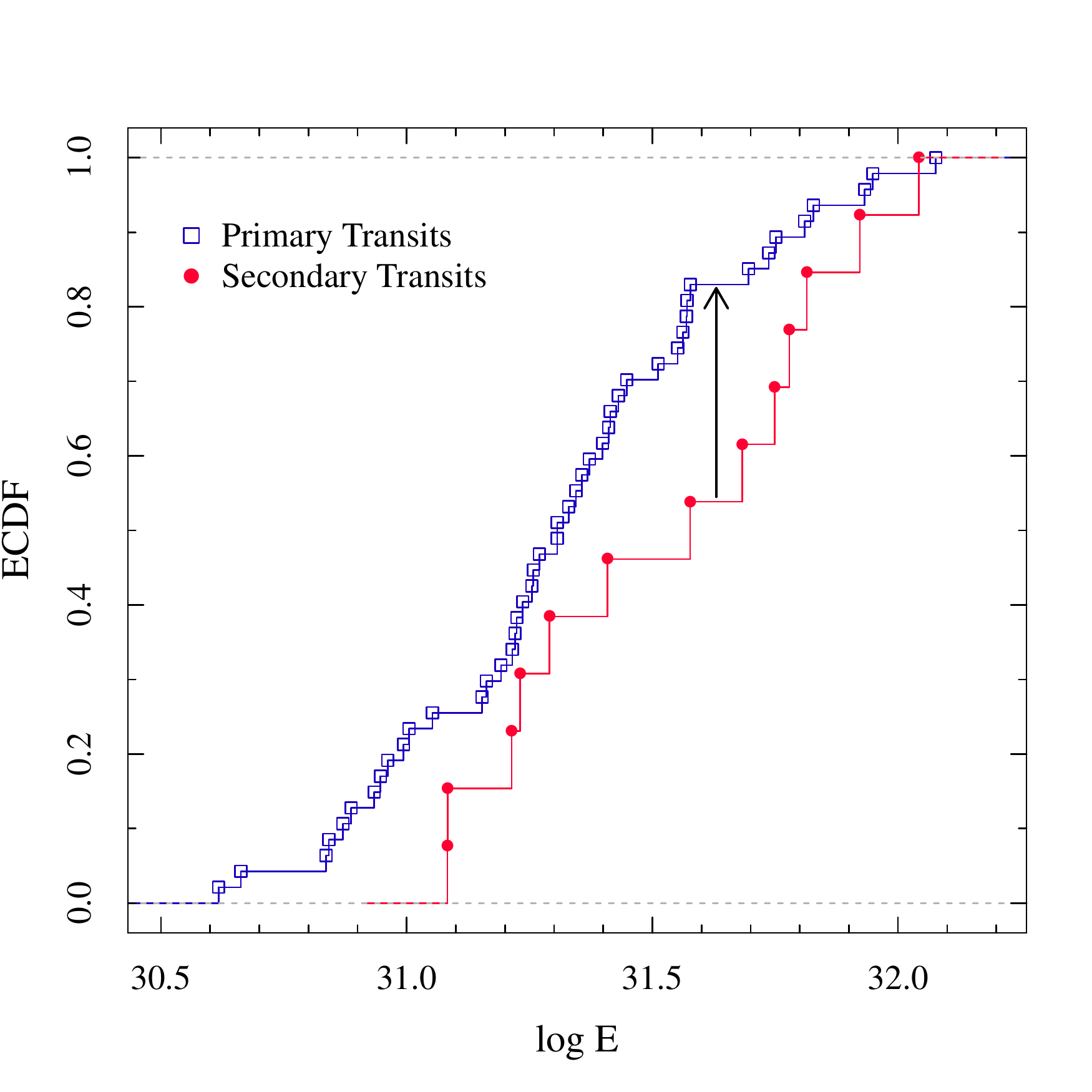} }
\end{center}
\caption{\label{fig:ecdf} Empirical cumulative distribution functions (ECDFs) 
of the energies of flares detected around primary and secondary transits. 
The arrow marks the largest separation between the two ECDFs at $\log E\approx 31.63$. }
\end{figure}

{ The range of flare energies spans about 1.2 orders of magnitude between $\log L_X \approx30.6$ 
erg and $\log L_X \approx 32.2$ erg. The medians of flare energies observed around primary 
and secondary transits are slightly different ($\log E_X\approx 31.31$  and 31.58, respectively) 
{ but the ranges of energies are similar in both sub-samples.}
A Kolmogorov-Smirnoff test applied to the two samples indicates a probability of $\sim13$\% 
that the energies of flares seen at the secondary transit are drawn from the same 
underlying distribution of energies of flares detected at the primary transits.
{ From this, we conclude that there is a slight, but not statistically significant 
hint that flares at the secondary transits were more energetic.
In Fig. \ref{fig:ecdf}, we show the  empirical cumulative distribution functions (ECDFs)
of the two samples of flare energies. The greatest difference between the two ECDFs is 
seen at $\log E\approx 31.6$.   } 
The peak luminosities and the duration of the flares of the two samples do not show a trend, 
meaning that it is likely a combination of flares duration and peak luminosities to result in the
{ slight} difference of energies. 

Figure \ref{logE} shows the rate of flares per day with energies above a given threshold 
for the total sample, the flares detected near the planetary transits and those detected 
near or after the secondary transits. 
The frequency of X-ray flares with respect to their energies in the Sun and in Main Sequence stars 
is well described by a power law relationship: \\

\indent $\frac{dN}{dE_X} \propto E_X^{-\alpha}$, \\

\noindent where the index $\alpha$ is of order of 2 \citep{Hudson1991,Kashyap2002,Drake2013}. 
The log-log distribution shown in Fig. \ref{logE} allows us to estimate the slope $\gamma$ 
of the linear relationship between the number of flares and
their energies: \\

\indent  $\log N(E>E_0) = \beta + \gamma \log E_X$,  \\

with the slope  $\gamma$ being related to $\alpha$ by the relationship $\gamma = 1 - \alpha$. 

The bias due to completeness of unidentified small flares makes the curves in Fig. \ref{logE} 
flatten below a given threshold ($E_{th}$). The value of $E_{th}$ depends on the sensitivity of the 
instruments and the intensity of the flares. Above $E_{th}$ 
the cumulative distribution  is deemed complete and not affected by unidentified smaller flares,
thus, it is possible to fit the tail of the distribution without any completeness bias.
We calculated the slope at different energy thresholds and checked that 
for values $\log E_X>31.2,$ the slope does not change significantly.
We thus considered only the 43/60 flares with energies above $\log E_{th} = 31.2$ that roughly 
correspond to flares  with peak luminosities between $5\times10^{27}$ and $2.8\times10^{28}$ \lxu 
or about one to six times the quiescent luminosities. 
In Sect. \ref{res}, we gave a statistical estimate of the number of small flares we could have missed. 
Their number amounts to about seven per light curve and 175 for the total observation time (11.1 days), 
or about ten small flares per day with energies $30.5< \log E<31.2$.  
However, in the case of a larger power law index $\alpha\sim-2.5,$ the number of undetected small 
flares with $30.5< \log E <31.2$  can be up to 40 per day.
{ In principle, our simulation of injected flares could be used to retrieve the completeness 
limit as a function of the flares energy. However, the flare model depends on a few choices we made
about the shape and the rise and decay characteristic times, which we kept fixed 
to reduce the space parameter. The flare energies were obtained from the integration of the 
flare profiles, but any given value of total energy could correspond to flares 
with different profiles and duration.
For this reason, we preferred to infer the completeness limit directly from the data.}
\begin{figure}
\resizebox{\columnwidth}{!}{
  \includegraphics{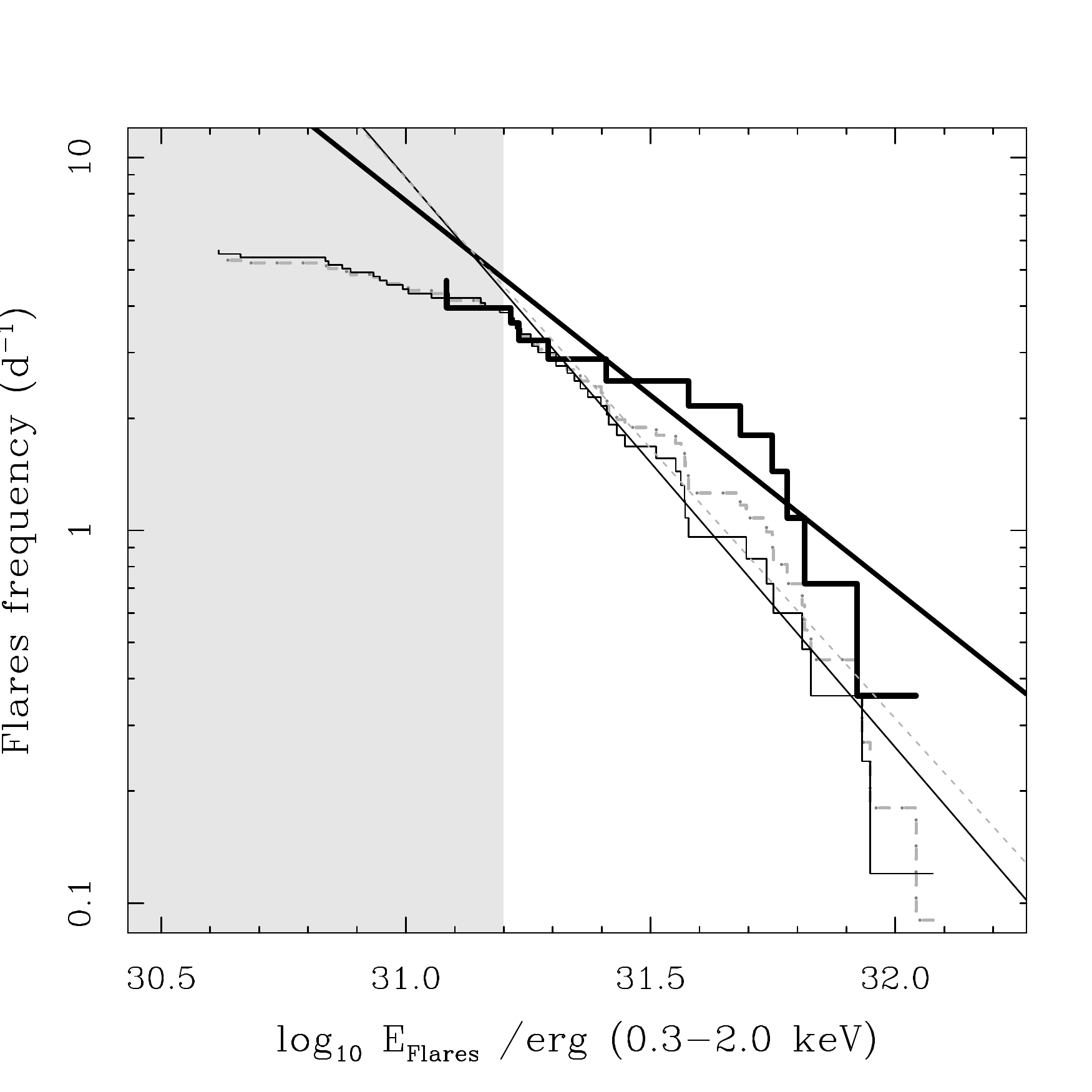}
}
\caption{\label{logE} Cumulative distributions of flares per day with energies above a given threshold 
($\log N(E>E_0)$). 
Dashed line indicates the full sample of flares ($N_\mathrm{flares}=60$); 
thin solid line is relative to the flares observed at planetary transits ($N_\mathrm{flares}=47$);
thick solid line is relative to the flares observed at secondary transits ($N_\mathrm{flares}=13$). 
The lines refer to a linear fit of each sample in the form $\log N(E_X>E_0) \propto \beta \log E_X$.
The range with energies $\log E_X<31.2$ not used for the fit to the slopes is shaded in gray color.
}
\end{figure}

For the  full sample of flares, we obtained: $\alpha \sim2.45\pm0.06$; 
for the flares observed around the primary transits: $\alpha \sim 2.53\pm0.05$; 
and for the flares observed at the secondary transit phases: $\alpha\sim2.04\pm0.14$.
Overall, the modeling of the distribution of flare energies with a power law is a good approximation,
however, the index of the power law that we find for \hd~A is steeper than the analog power 
law indices found is the Sun and in other solar-type stars. 

The subsample of flares detected at the secondary transits shows a flatter energy distribution 
than the corresponding distribution at the primary transits. 
The respective power law indices differs by about $2-3\sigma$ { , although 
the uncertainties in the slopes could be underestimated given the correlation of the errors among
the  bins.}

We go on to consider whether the flares observed in \hd~A could be associated with coronal mass ejections (CMEs).
For the Sun, \citet{Yashiro2004} observed that flares of classes higher than X2 (corresponding 
to fluxes of 0.2 \fxu) are always associated with CMEs.  
We calculated the average luminosities ($0.3-2.0$ keV) during flares and converted 
these to the GOES band ($1-8\AA$), with a  scaling factor
calculated assuming an APEC model with $kT=0.86$ keV, which is appropriate for 
the peak temperatures we inferred from HR, assuming $N_H=0$ cm$^{-2}$ and $Z=Z_\odot$. 
The scaling factor was thus 0.1122. 
From the luminosities, we derived the fluxes at 1 AU and compared these to the GOES scale. 
Using Yashiro's classification, a hypothetical Earth around \hd~A would have experienced 
flares of classes between X1 and X20, more than 70\% of them associated with CMEs. 

\citet{Drake2013} give an estimate of the mass rate of CMEs as a function of the GOES  
luminosity. The average luminosities of the flares of \hd~A converted in the GOES band 
give values above $6\times10^{28}$ erg. From Fig. 3 in \citet{Drake2013}, we infer 
that the mass rates ejected in the supposed CMEs are in excess of $10^{-12} M_\odot yr^{-1}$,
with most of them peaking at $10^{-11} M_\odot$ yr$^{-1}$ and reaching up to $10^{-10} M_\odot yr^{-1}$
in the most powerful flares, or about $6-600 \times 10^{13}$ g s$^{-1}$.
\citet{Drake2013} give also the relationship between X-ray luminosity of flares and kinetic energy 
loss rate of the CME. From the flare luminosities of \hd~A, we infer a kinetic energy in excess of 
$10^{-4}$ L$_\odot$ and a velocity of $v_{CME}\approx 1000$ km s$^{-1}$.

Since the average X-ray activity of the stars declines with age, it is expected that also the most 
energetic flares  become less frequent (see \citealp{Johnstone2021}). 
\citet{Audard2000} derived an empirical relationship between the number of flares per day with 
energies above $10^{32}$ erg and the average X-ray luminosities of the stars in the band 0.01--10 keV. 
After converting the luminosities in 0.0--10 keV, we detected five flares with energies above $10^{32}$ erg 
in 11.1 days or N(E>$10^{32}\mathrm{erg})\approx0.45\, \mathrm d^{-1}$  flares. 
The relationship described by Audard et al. (2000) for an average luminosity of \hd~A of $6\times10^{28}$ \lxu predicts
about 0.46 flares per day, which is perfectly matched with the observed rate.
Two out of the five flares with E>$10^{32}$ erg were detected at secondary transits, 
again hinting that energetic flares occur preferentially at those phases.
The finding that more energetic flares occur more frequently after the planetary eclipse is in agreement with
prior results \citep{Pillitteri2011,Pillitteri2015}. 
The reason for this could be some kind of star-planet interaction due to the escaping planetary gas 
trapped in the global magnetic field around the star. The interplay between ionized gas 
and magnetic field could generate reconnection events leading to energetic flares (cf. \citealp{Cohen2011}). 
In support of this scenario, the analysis of the flare observed in ObsID 0690890201 \citep{Pillitteri2014} 
led us to conclude that the flaring structure had a size of a few stellar radii, which is compatible 
with a magnetic structure elongated towards the planet -- produced by the dynamics of the ionized gas escaping 
the planet and launched towards the star (Colombo et al. 2021 in prep.,  see also the MHD modeling 
in \citealp{Matsakos2015}).
 Using Eq. 14 of \citet{Serio1991} and by assuming a typical decay time $\tau_d= 5$ ks and a peak 
temperature of 0.9 keV, we estimated an average loop length of $\approx1.6$ stellar radii. For longer lasting 
flares with $\tau_d \sim10$ ks, the loop would be twice in length with the top of the loop distant about 1 stellar
radius above the stellar surface. 
{ \hd\  was one of the first cases for which star-planet interaction was suggested from
the varying emission of Ca II H\&K lines \citep{Shkolnik2008}. 
\citet{Pillitteri2014} derived a loop length of about four stellar radii from the 
analysis of an oscillatory pattern detected in a flare light curve. 
The height of this loop was close to 25\% of the separation between the planet and the star 
and it is above the star’s co-rotation radius. The latter indicates there is a mechanism that 
maintained the loop stable. The authors suggested that the star and the planet are magnetically 
connected. This hypothesis was later supported by MHD simulations and FUV observations 
\citep{Matsakos2015,Pillitteri2015}. The latter authors speculated that the star could accrete 
gas evaporating from the planet. The impact of the stream onto the stellar surface would cause 
a hot spot that is approximately 90 degrees ahead of the sub-planetary point. 
Pillitteri et al. (2015) also showed that the star exhibited FUV variability in phase with the 
planet motion. The hot spot would become visible during the secondary transit and be hidden to 
the observer during the primary transit. 
The stellar activity in the hot spot is higher in terms of flare occurrence and intensity but 
does not cause a perceptible enhancement in the quiescent emission of the star, 
possibly due to the reduced extension of the spot. Further MHD simulations show that the
stream from the planet to the star is dynamically varying due to the interaction 
with the stellar wind \citep{Colombo2021}. 
In a concurrent scenario (not necessarily an alternative to the first one), 
the magnetized environment between the planet and the star traps the out-flowing ionized gas from 
\hd Ab, but the planetary orbital motion stretches the resulting magnetic field and 
triggers the flares (cf. \citealp{Lanza2013}). 
These speculations are, however, biased by the fact that we have observed a relatively narrow portion of
planetary orbital phases and that a continuous and repeated monitoring covering 
at least half orbit between the two opposite phases has not been obtained. 
} 
Further monitoring at the secondary transit phases is desirable to match the unbalanced exposure times
at the two opposite phases and to better determine any difference of 
flare energies  when the planet is facing the two different sides of the star.  

The average temperature of the corona of \hd A is estimated at around 0.4 keV (10\%-90\% range in 0.27-0.58 keV). 
It is only during flares that the temperature rises up to about 0.9 keV. 
The flares observed near primary transits and those observed at secondary transits show similar temperatures.
The lack of temperatures in excess of 1 keV is not due to a lack of sensitivity
of our analysis (which can, in principle, detect temperatures up to $\sim1.5$ keV). 
The corona of \hd~A remained relatively cold even during the most energetic flares, while, for example, 
\hd~B showed temperatures on the order of $1.3-1.5$ keV during the four flares that we observed and that 
released energies $E \approx 10^{32}$ erg. 

From the monitoring of chromospheric lines as activity indicators and in X-ray,
a number of stars are known to possess activity cycles even among planet host stars. 
The ubiquity of such cycles are not established and depends on several factors such as the stellar age. 
\citet{Saar1999} and \citet{BohmVitense2007} showed that when comparing rotational periods with the 
lengths of activity cycles, stars can be divided into two sequences: a steeper sequence formed by young and 
active stars and a flatter sequence of inactive and older stars. The Sun fits in between the two sequences. 
\hd~A does not seem to possess an X-ray activity cycle, at least when considering a timescale of eight years. 
\hd~A has a rotational period of 11.7 days, as an active star (mimicking an
age of less than 1 Gyr), it belongs to the steep sequence of active and young stars 
and, thus, it would have an activity cycle of $\geq12$ years. 

{ One peculiar behavior characterizing \hd~A is the fact that its variability in X-rays is totally different 
from the variability in the optical band. 
TESS observed \hd~A (TIC~256364928) during sector 41 for about 27 days and about 11 transits
were recorded. The TESS light curve does not clearly show short timescale variability due to flares. 
As remarked in Sect. \ref{res}, the planet obscured some surface spots but yet the star was 
relatively inactive, behaving instead like an old quiet star in spite of its level of X-ray activity and luminosity.
We speculate that the stark difference between the activity and the relatively cold flares observed in
X-rays and the optical band could be related to the angular momentum transfer from the 
planet to the star and its influence on the mechanism of magnetic dynamo that produces the corona.
Furthermore, the magnetic interaction between the star and the planet could produce a significant part
of the observed X-ray emission and variability. \citet{Lanza2013} quantifies the power emitted by
two possible mechanisms of loop stretching due to magnetic interaction of the star+planet magnetospheres
or of magnetic loops anchored between a star and its hot Jupiter. 
In particular, the second mechanism could easily produce a power of order of $10^{28}$ \lxu 
(similar to the average X-ray luminosity of \hd~A) for a planetary magnetic field of
$\sim3$ G at the pole and separation equal to that of \hd~Ab. 
} 

We analyzed the variability of \hd~B, taking into account the close angular separation 
with the primary \hd~A  in order to carefully distinguish its time variability in X-ray. 
This mainly consisted of four bright flares detected in 958 ks,
which correspond to a rate of one flare every 2.8 days. 
On average, the flares released energies in excess of $10^{32}$ erg.
The quiescent luminosity of \hd~B is about $\log L_X\sim 27.27$ in $0.3-2.0$ keV, 
which is about a factor four higher than found in Chandra observations \citep{Poppenhaeger2013}.  
The luminosity is still consistent with that of  a dM star with an age of a  it is Gyr; however, in contrast 
to the age that would otherwise be inferred from the X-ray activity 
of \hd~A, which is less than 1 Gyr. The age discrepancy between the two components of the HD 189733 system 
has been discussed in literature and explained as the result of spin up on the part of the star with the hot Jupiter, accompanied by the transfer 
of angular momentum from the orbital motion of the planet to the rotation of the star 
\citep{Pillitteri2010,Poppenhaeger2014}. 
The flare temperatures of \hd~B were in excess of 1.3 keV -- in stark contrast to the flare temperatures
of the flares of \hd~A, which remained below 1 keV.

\begin{acknowledgements}
The authors acknowledge the anonymous referee for their useful comments and suggestions. 
IP, AM and GM acknowledge financial support from the ASI-INAF agreement n.2018-16-HH.0, and from 
the ARIEL ASI-INAF agreement n.2021-5-HH.0. Based on observations obtained with XMM-Newton, 
an ESA science mission with instruments and contributions directly funded by ESA Member States and NASA 
\end{acknowledgements}


\begin{appendix}
\section{Table of flares}
\onecolumn
\begin{table}
\caption{\label{fltab}  List of identified flares. }
\resizebox{0.99\textwidth}{!}{
\begin{tabular}{cccccc|cccccc}\\\hline \hline
Nr. Obs. & Duration      & $\log E_X$    & $\log L_X$    & $\log f_X$    &$\log f_{X,Pl}$& Nr. Obs.    & Duration & $\log E_X$    & $\log L_X$    & $\log f_X$    & $\log f_{X,Pl}$ \\ 
         &      ks       &      erg      &       \lxu    &      \fxu     &      \fxu     &               &       ks       &      erg     &      \lxu     &      \fxu    &      \fxu  \\\hline
1        &      2.4      &      30.89    &      28.11    &      -12.56   &       3.68     &      13       &      5.4      &      31.24    &      28.46    &       -12.21   &      4.03  \\
1        &      3.6      &      31.00    &      28.23    &      -12.44   &       3.79     &      13       &      4.2      &      31.26    &      28.48    &       -12.19   &      4.05  \\
1        &      6        &      31.21    &      28.44    &      -12.23   &       4.00     &      14       &      4.8      &      31.15    &      28.38    &       -12.3    &      3.94  \\
2        &      18.6     &      32.04    &      29.26    &      -11.41   &       4.83     &      15       &      8.4      &      31.51    &      28.73    &       -11.94   &      4.3  \\
3        &      7.8      &      31.75    &      28.97    &      -11.7    &       4.54     &      16       &      6.0      &      31.58    &      28.80    &       -11.87   &      4.37  \\
3        &      6.6      &      31.41    &      28.63    &      -12.04   &       4.20     &      16       &      13.2     &      31.74    &      28.96    &       -11.71   &      4.53  \\
4        &      6        &      31.21    &      28.44    &      -12.23   &       4.00     &      16       &      5.4      &      31.43    &      28.65    &       -12.02   &      4.22  \\
4        &      8.4      &      31.29    &      28.51    &      -12.16   &       4.08     &      17       &      10.8     &      31.75    &      28.97    &       -11.7    &      4.54  \\
4        &      22.8     &      31.92    &      29.14    &      -11.53   &       4.71     &      19       &      5.4      &      31.36    &      28.58    &       -12.09   &      4.15  \\
5        &      9        &      31.57    &      28.79    &      -11.88   &       4.36     &      19       &      1.8      &      30.62    &      27.84    &       -12.83   &      3.41  \\
5        &      5.4      &      31.40    &      28.62    &      -12.05   &       4.19     &      20       &      3.0      &      30.84    &      28.06    &       -12.61   &      3.63  \\
5        &      7.8      &      31.16    &      28.38    &      -12.29   &       3.95     &      20       &      3.0      &      30.87    &      28.09    &       -12.58   &      3.66  \\
5        &      7.8      &      31.41    &      28.63    &      -12.04   &       4.20     &      20       &      14.4     &      31.81    &      29.03    &       -11.64   &      4.6  \\
6        &      2.4      &      30.96    &      28.18    &      -12.49   &       3.75     &      21       &      10.2     &      31.83    &      29.05    &       -11.62   &      4.62  \\
7        &      4.8      &      31.33    &      28.55    &      -12.12   &       4.12     &      21       &      8.4      &      31.56    &      28.78    &       -11.89   &      4.35  \\
7        &      2.4      &      30.95    &      28.17    &      -12.5    &       3.74     &      21       &      5.4      &      31.22    &      28.44    &       -12.23   &      4.01  \\
7        &      1.8      &      30.66    &      27.88    &      -12.79   &       3.45     &      22       &      6.0      &      31.70    &      28.92    &       -11.75   &      4.49  \\
7        &      5.4      &      31.41    &      28.64    &      -12.03   &       4.20     &      22       &      18.0     &      31.93    &      29.15    &       -11.52   &      4.72  \\
8        &      6        &      31.31    &      28.53    &      -12.14   &       4.10     &      23       &      5.4      &      31.19    &      28.41    &       -12.26   &      3.98  \\
8        &      8.4      &      31.55    &      28.77    &      -11.9    &       4.34     &      23       &      4.2      &      31.25    &      28.48    &       -12.19   &      4.04  \\
9        &      2.4      &      30.99    &      28.22    &      -12.45   &       3.78     &      23       &      2.4      &      31.05    &      28.27    &       -12.4    &      3.84  \\
9        &      6        &      31.31    &      28.53    &      -12.14   &       4.10     &      23       &      6.0      &      31.57    &      28.79    &       -11.88   &      4.36  \\
9        &      2.4      &      30.84    &      28.06    &      -12.61   &       3.63     &      23       &      2.4      &      30.93    &      28.16    &       -12.51   &      3.72  \\
10       &      6.6      &      31.45    &      28.67    &      -12      &       4.24     &      24       &      6.0      &      31.78    &      29.00    &       -11.67   &      4.57  \\
10       &      4.2      &      31.22    &      28.45    &      -12.22   &       4.01     &      24       &      8.4      &      31.68    &      28.90    &       -11.77   &      4.47  \\
10       &      6        &      31.37    &      28.59    &      -12.08   &       4.16     &      24       &      3.0      &      31.08    &      28.30    &       -12.37   &      3.87  \\
11       &      4.8      &      31.34    &      28.57    &      -12.1    &       4.13     &      24       &      10.8     &      31.58    &      28.80    &       -11.87   &      4.37  \\
11       &      4.8      &      31.27    &      28.49    &      -12.18   &       4.06     &      25       &      4.2      &      31.23    &      28.45    &       -12.22   &      4.02  \\
11       &      13.2     &      31.95    &      29.17    &      -11.5    &       4.74     &      25       &      15.0     &      31.81    &      29.04    &       -11.63   &      4.6  \\
12       &      22.2     &      32.08    &      29.30    &      -11.37   &       4.87     &      25       &      6.0      &      31.08    &      28.31    &       -12.36   &      3.87  \\
\hline
\end{tabular}\\
}
Note:  We report the observation number (cf. Table \ref{obslog}), 
the duration of the flares, their  energy in the band 0.3-2.0 keV, the total emitted luminosities and 
fluxes in the same band, and the fluxes received at the mean separation of \hd~Ab. 
\end{table}

\section{Light curves}
In this appendix, we display the \pn\ light curves of \hd~A (Fig. \ref{lc1}) obtained in 25 \xmm\ observations,
the HR light curves derived from the \pn\ count rates in the bands $0.3-1.0$ keV and  $1.0-2.0$ keV (Fig. \ref{hrpn}), 
and the light curves of the X-ray luminosity of \hd~A in the band $0.3-2.0$ keV (Fig. \ref{lxall}).

\begin{figure*}
\resizebox{0.99\textwidth}{!}{
\includegraphics{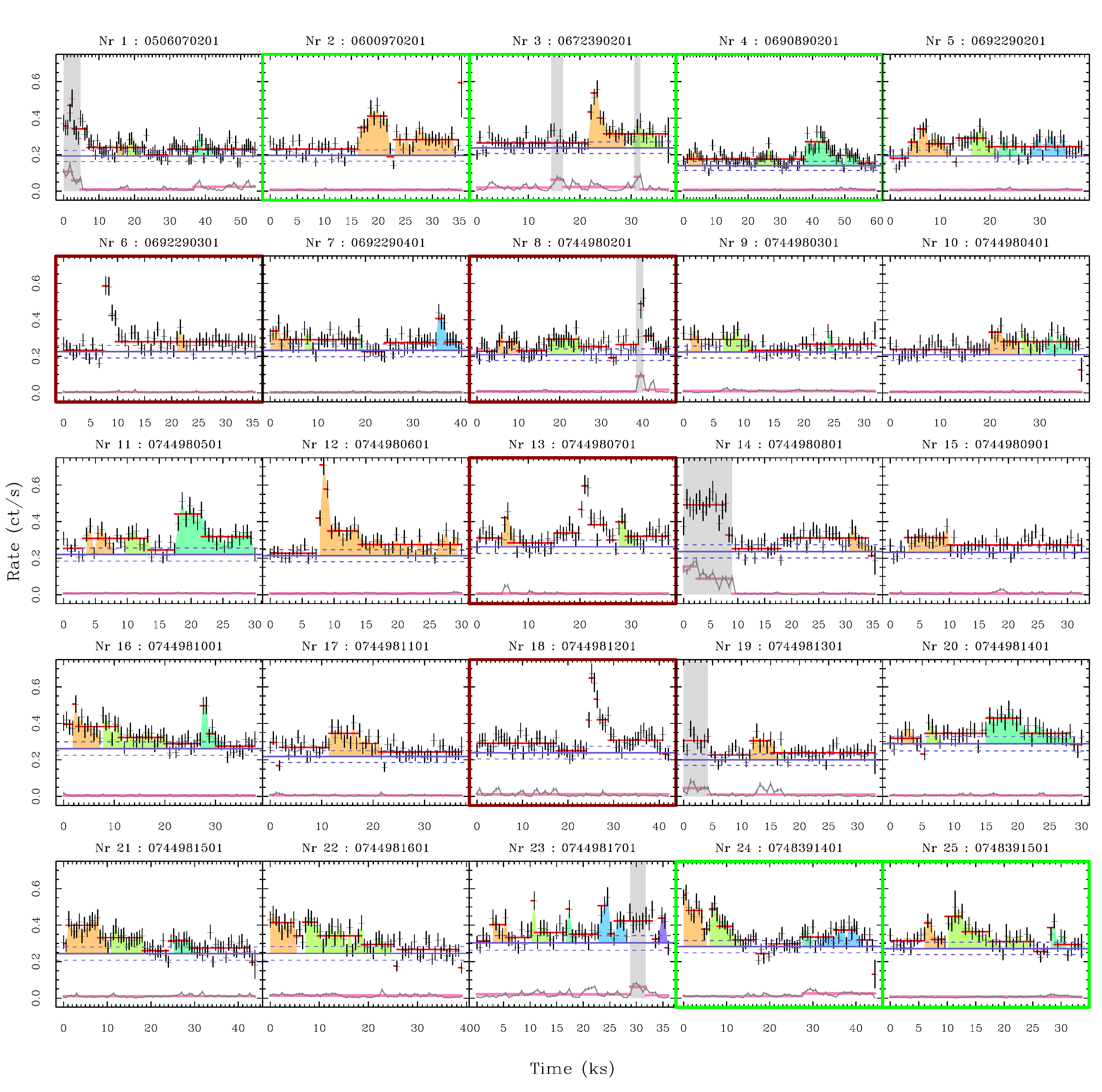}
}
\caption{\label{lc1}
Light curves of \hd~A obtained with \pn\ in the band $0.3-2.0$ keV. 
The points with error bars are the net count rates of \hd~A for the events in the selection
operated at points {\em i} and {\em ii} of Sect. \ref{obs}. 
The horizontal red segments mark the intervals identified with the {\em Changepoint} algorithm.
Similarly, the light red segments represent the intervals of constant background. 
In intervals of high background variability (gray shaded areas), we did not search for
flares in \hd~A. The flares are shaded with different colors in each panel.
The horizontal solid blue lines represent the quiescent rates and the $\pm1sigma$ range.
Panels contoured with a green border refers to observations at the secondary transits. Panels marked with a red border indicate where variability of \hd~B was prominent.}
\end{figure*}
\begin{figure*}
\resizebox{0.99\textwidth}{!}{
\includegraphics{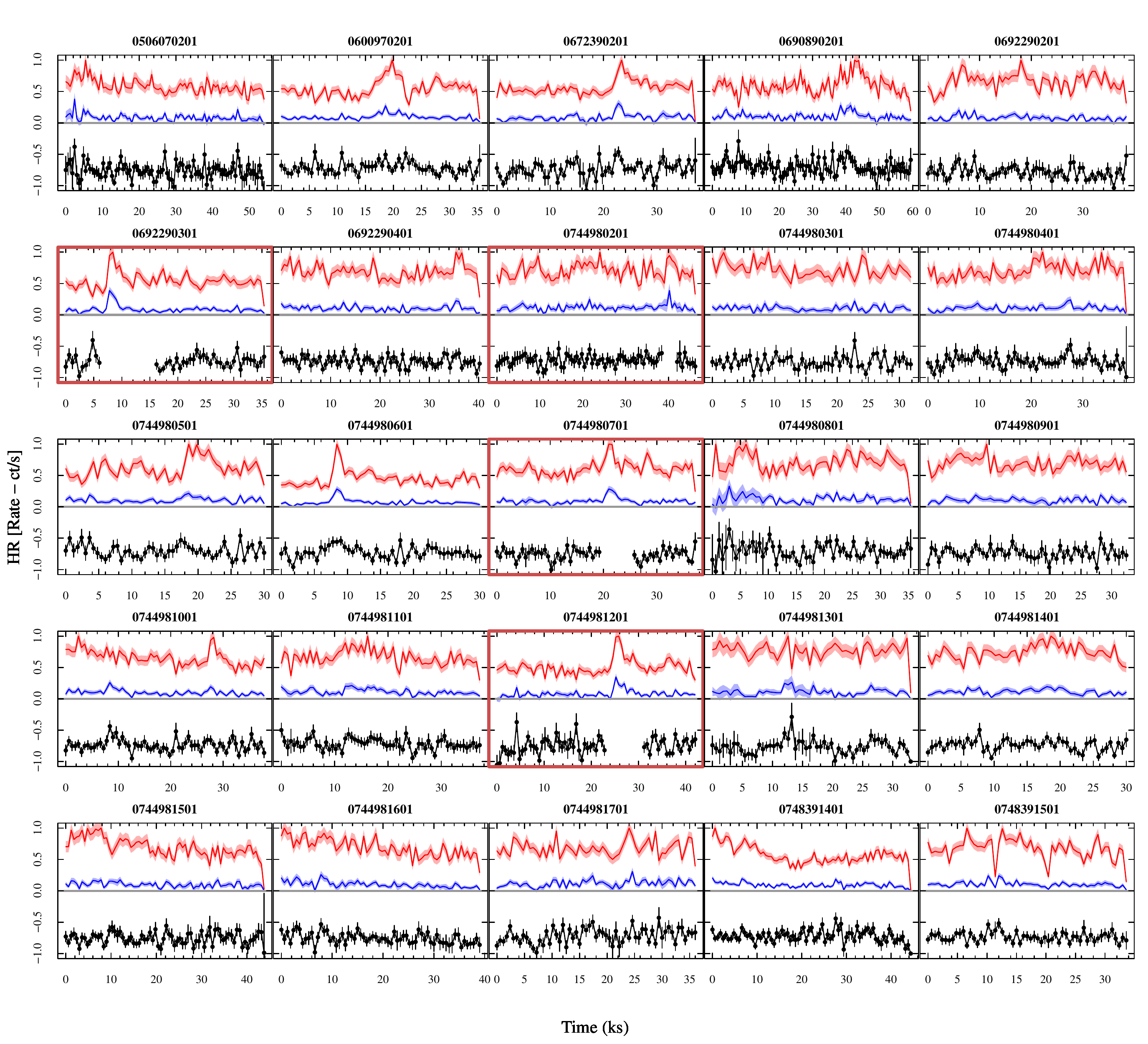}
}
\caption{\label{hrpn}
Solid  black lines with error bars: 
light curves of hardness ratio of  \hd~A obtained from \pn\ events for each observation. 
Red and blue curves: Soft (0.3-1.0 keV) and hard (1.0-2.0) band light curves scaled between
zero and the maximum of soft rate. 
As in Fig. \ref{lc1}, the panels marked with a red border indicate observations where flares of \hd~B were detected.
In these curves, we discarded the intervals relative to the flares of \hd~B.}
\end{figure*}
\begin{figure*}
\resizebox{0.99\textwidth}{!}{
\includegraphics{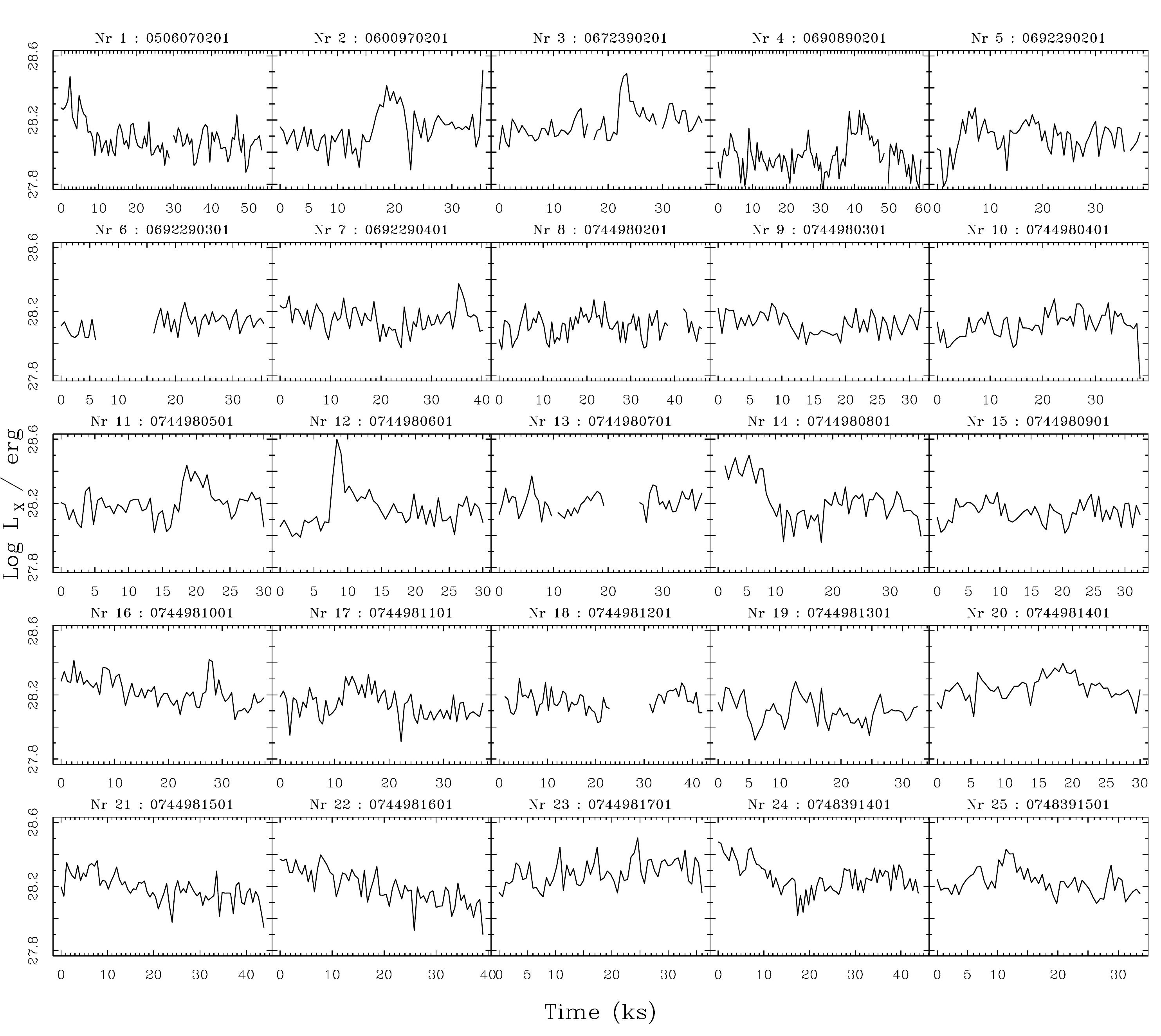}
}
\caption{\label{lxall}
Synoptic plot of the logarithm of the X-ray luminosity of \hd~A in the band 0.3-2.0 keV observed in the \xmm\ observations. Gaps in the 
light curves correspond to intervals where flares from \hd~B affected the estimated of the count rate of \hd~A.}
\end{figure*}
\end{appendix}
\end{document}